\documentclass{article}
\usepackage{spconf,amsmath,graphicx,float,caption,hhline,fancyhdr}
\usepackage[hyphens]{url}
\usepackage[colorlinks,citecolor=magenta,urlcolor=blue,bookmarks=false,hypertexnames=true,linkcolor=cyan]{hyperref} 
\usepackage[hyphenbreaks]{breakurl}
\floatstyle{plaintop}
\restylefloat{table}
\pdfoutput=1

\fancypagestyle{copyrightfooter}{
  \fancyhf{}
  
  \fancyfoot[C]{\scriptsize {\copyright 2021 IEEE. Personal use of this material is permitted. Permission from IEEE must be obtained for all other uses, in any current or future media, including reprinting/republishing this material for advertising or promotional purposes, creating new collective works, for resale or redistribution to servers or lists, or reuse of any copyrighted component of this work in other works.}}
}

\title{Deep Image Debanding}

\name{Raymond Zhou, Shahrukh Athar, Zhongling Wang, and Zhou Wang}
\address{Department of Electrical \& Computer Engineering, University of Waterloo, Canada \\ Email: \{raymond.zhou, shahrukh.athar, zhongling.wang, zhou.wang\}@uwaterloo.ca \vspace{3mm}}

\begin{document}
\ninept
\maketitle

\begin{abstract}
Banding or false contour is an annoying visual artifact whose impact is even more pronounced in ultra high definition, high dynamic range, and wide colour gamut visual content, which is becoming increasingly popular. Since users associate a heightened expectation of quality with such content and banding leads to deteriorated visual quality-of-experience, the area of banding removal or debanding has taken paramount importance. Existing debanding approaches are mostly knowledge-driven. Despite the widespread success of deep learning in other areas of image processing and computer vision, data-driven debanding approaches remain surprisingly missing. In this work, we make one of the first attempts to develop a deep learning based banding artifact removal method for images and name it \textit{deep debanding network} (deepDeband). For its training, we construct a large-scale dataset of 51,490 pairs of corresponding pristine and banded image patches. Performance evaluation shows that deepDeband is successful at greatly reducing banding artifacts in images, outperforming existing methods both quantitatively and visually. 

\thispagestyle{copyrightfooter}


\end{abstract}


\section{Introduction}
\label{sec:intro}

\begin{figure}[t!]
    \begin{minipage}[b]{1.0\linewidth}
        \centering
        \centerline{\includegraphics[width=8.5cm]{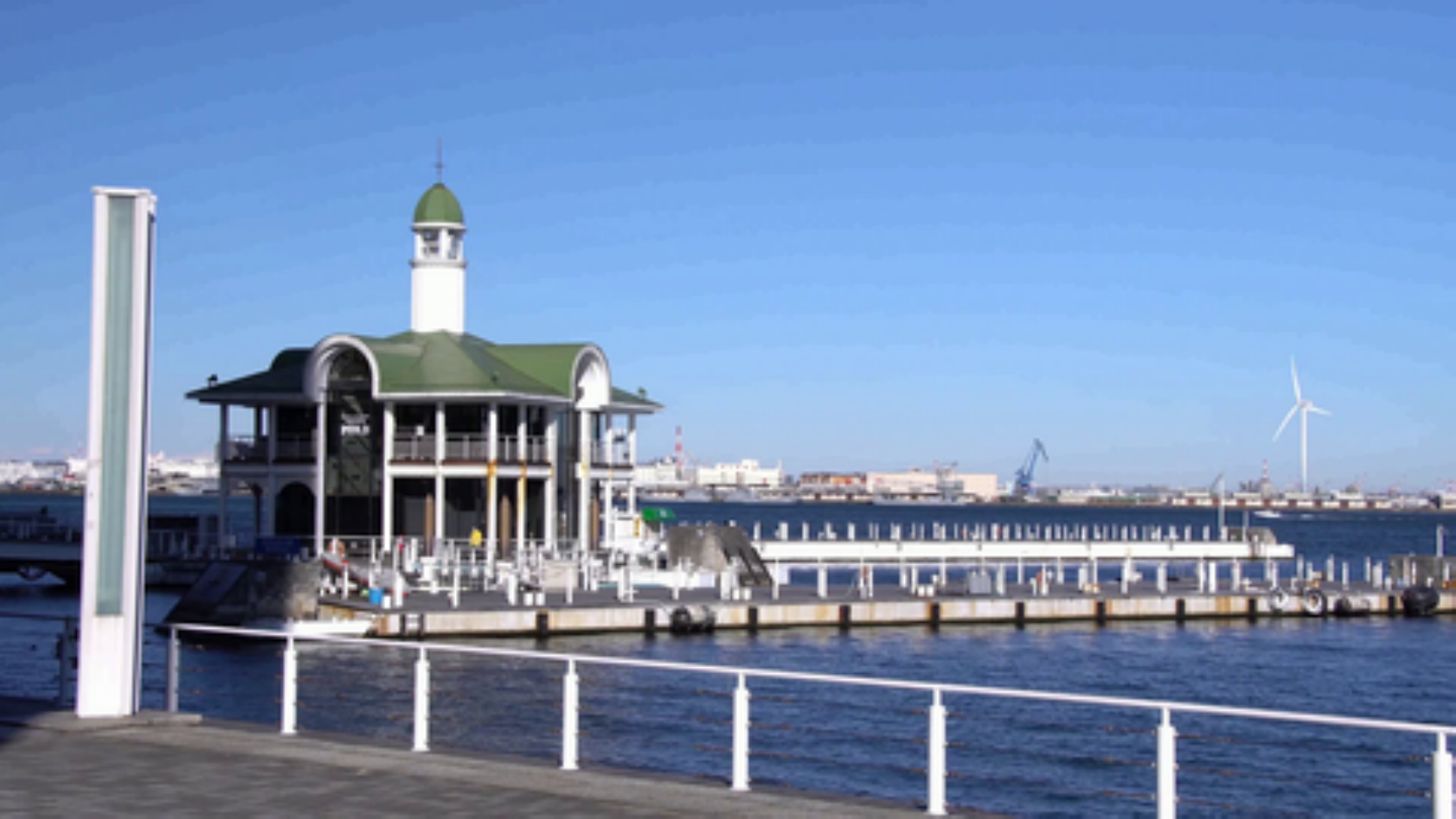}}
    \end{minipage}
    \caption{Example of banding artifacts present in the sky region.}
    \label{fig:example}
    \vspace{-3mm}
\end{figure}

Banding or false contour artifacts are common visual annoyances found in visual content and are caused by quantization. They often occur in large regions of smooth visual content with low textures and slow gradients, such as sky or water. Banding manifests as sharp, discrete colour discontinuities where there otherwise should be smooth transitions, causing notable degradation in visual quality. Fig. \ref{fig:example} shows an example of an image with severe banding artifacts visible in the sky region. Recent technological advances have led to the popularity and widespread adoption of ultra high definition (UHD), high dynamic range (HDR), and wide colour gamut (WCG) visual content, where users typically expect better visual quality-of-experience. While banding artifacts have been studied in standard dynamic range (SDR) content, their analysis remains limited for HDR content \cite{hdr}. Visually, banding artifacts appear even more annoying in UHD/HDR/WCG content, which is counter-intuitive as one would expect that increasing bit-depth would resolve this problem. However, UHD/HDR/WCG content covers a much wider range of luma and chroma levels, compared to SDR, which exacerbates the banding problem. Thus, there is an urgent need to develop accurate banding detection and banding removal (or debanding) methods that are practically applicable.


Contemporary debanding methods such as FFmpeg's deband filter \cite{ffmpeg}, AdaDeband \cite{adadeband}, and FCDR \cite{fcdr} are knowledge-driven, i.e., they rely on particular domain knowledge and understanding of the human visual system \cite{ibc}. A major disadvantage of such methods is that they have multiple parameters that must be calibrated properly for optimal results, which can be a lengthy and challenging process. Additionally, many knowledge-driven methods like \cite{fcdr} employ dithering, which introduces noise, as a means to lessen the visibility of banding. However, this process often reduces the visibility of fine texture details, which is especially a problem for UHD/HDR/WCG content that is meant to present finer details \cite{ibc}. An alternative is to take a data-driven approach where machine learning, especially deep learning, is employed to learn required models. Indeed, deep learning techniques have been widely used in the image restoration context, for example, in the removal of noise \cite{noise}, blur \cite{blur}, and blocking artifacts \cite{blocking}. However, to the best of our knowledge, thus far there has been no effort to use deep learning targeted specifically at banding removal. 

In this work, we make one of the first attempts to develop a deep learning based model for removing banding artifacts from images, taking a banded image and returning its debanded version, and call it \textit{deepDeband}. Since a major bottleneck in the development of robust deep learning models is a lack of annotated training data, we construct a large dataset of 51,490 pairs of image patches with and without banding artifacts, as explained in Section \ref{sec:data}, and use it to train deepDeband. In Section \ref{sec:dev}, we present the architecture of the neural network, two different ways of applying the model to the input image, and development details. Finally, we present a quantitative and visual analysis of deepDeband's performance in Section \ref{sec:comp}, demonstrating its success over existing debanding methods.

\section{Dataset Construction}
\label{sec:data}

To construct a new dataset that enables training of deep learning based debanding models, we start with an existing dataset \cite{dbi}, which contains 1,439 pairs of pristine and their corresponding quantized images of 1920$\times$1080 resolution (FHD), where the quantized images have been segmented and labelled into banded and non-banded regions. To the best of our knowledge, this is the only publicly available dataset containing labelled banded images of this kind. From each quantized FHD image, we extract overlapping image patches of size 256$\times$256 with a sliding window of stride 75. Using the image labels from \cite{dbi}, we select only those patches that contain banding and also extract their corresponding patches from the pristine FHD images. This process results in 51,490 pairs of image patches, which are partitioned into training ($\sim$60\%), validation ($\sim$20\%), and test ($\sim$20\%) sets without content-overlapping, such that for any FHD image, all patches extracted from it belong to the same set. Each set contains images of diverse visual content, the scale of which allows for developing deep learning models. 

\begin{figure}[t!]
    \begin{minipage}[b]{1.0\linewidth}
        \centering
        \centerline{\includegraphics[width=8.5cm]{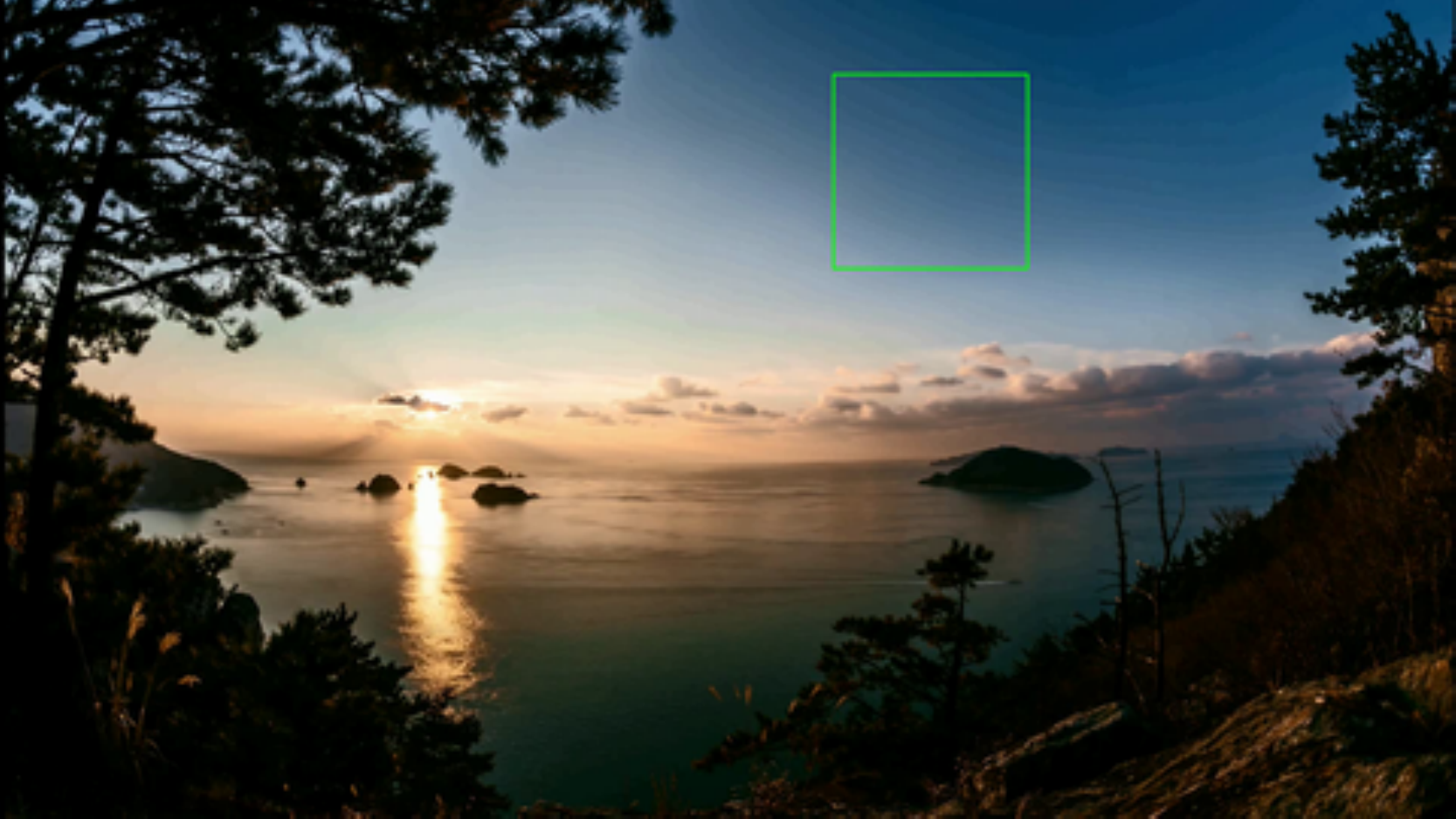}}
        \centerline{(a) Banded FHD image}\medskip
    \end{minipage}
    \begin{minipage}[b]{0.48\linewidth}
        \centering
        \centerline{\includegraphics[width=4.0cm]{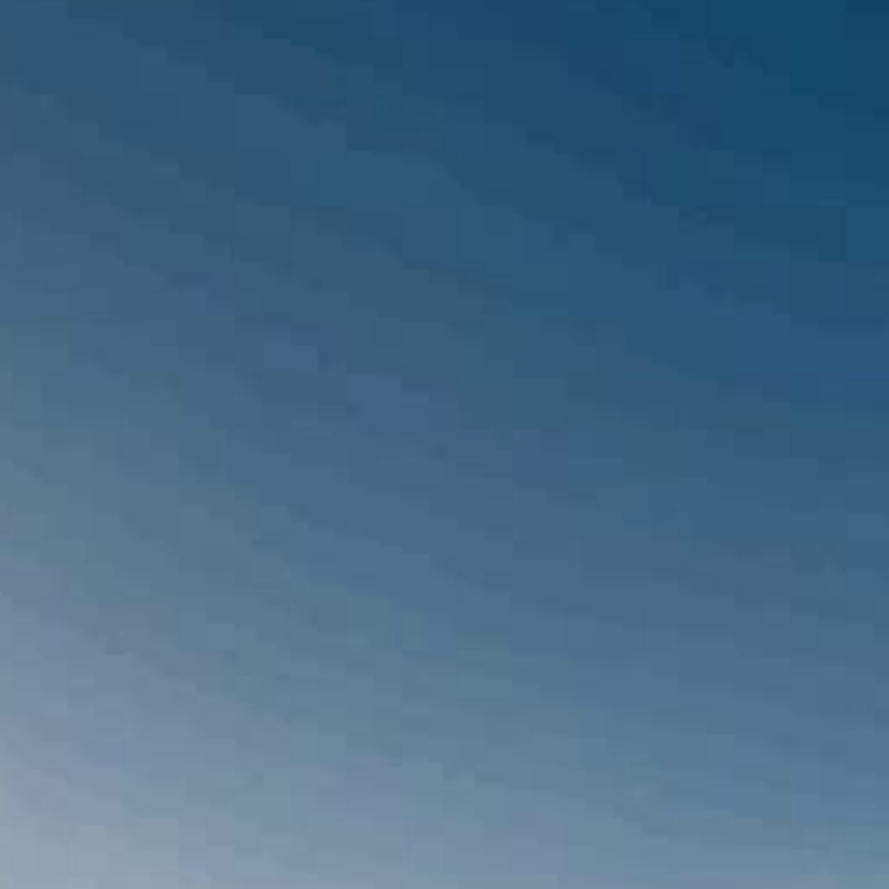}}
    \centerline{(b) Banded patch}\medskip
    \end{minipage}
    \hfill
    \begin{minipage}[b]{0.48\linewidth}
        \centering
        \centerline{\includegraphics[width=4.0cm]{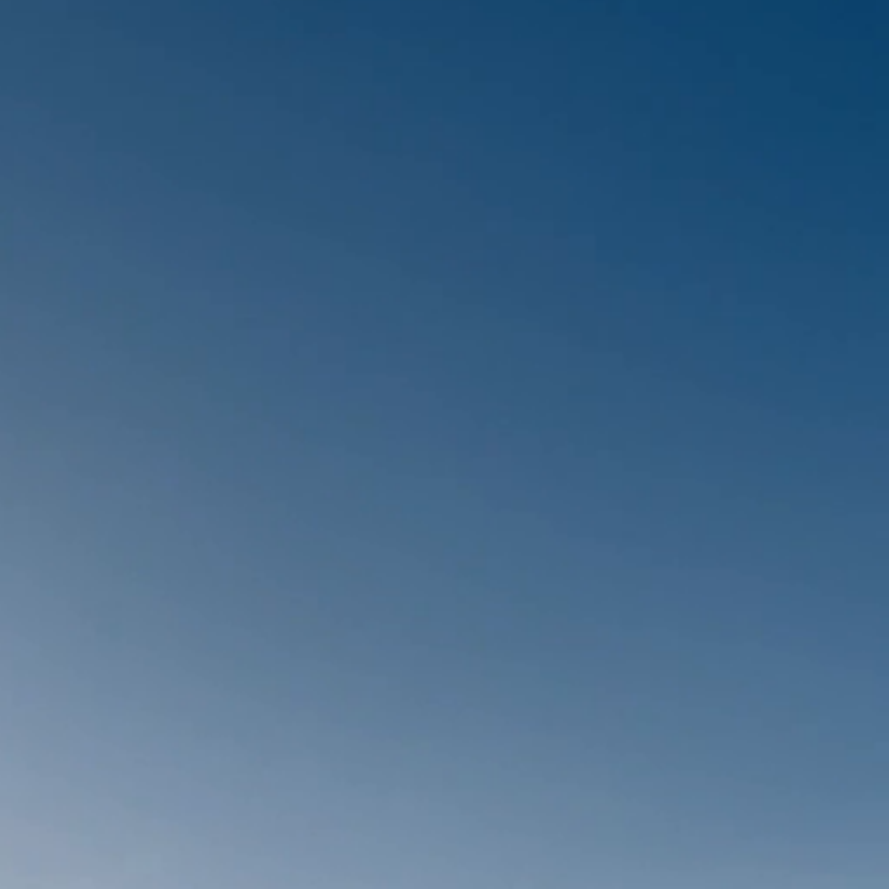}}
        \centerline{(c) Pristine patch}\medskip
    \end{minipage}
    \caption{Banded FHD image (a), extracted banded patch (b), and corresponding pristine patch extracted from the pristine FHD image (c).}
    \label{fig:patches}
    \vspace{-4mm}
\end{figure}

Fig. \ref{fig:patches} gives an example of a banded FHD image with a zoomed-in banded patch and its corresponding pristine patch that has been extracted from the respective pristine FHD image. Table \ref{tab:dataset} provides a detailed overview of the dataset. Since different FHD images can result in fairly different numbers of banded patches, the percentage of FHD images corresponding to each of the training, validation, and test sets is not in the same proportion as that of the image patches.



\begin{table}[t!]
    \small
    \centering
    \begin{tabular}{ c|c|c } 
    \hline
    Dataset & Patches (256$\times$256) & FHD (1920$\times$1080) \\ \hline \hline
    Training & 30,988 & 872 \\ \hline
    Validation & 10,203 & 257 \\ \hline
    Testing & 10,299 & 310 \\ \hline \hline
    Total & 51,490 & 1439 \\ \hline
    \end{tabular}
    \caption{Composition of the dataset in terms of patches and FHD images.}
    \label{tab:dataset}
    \vspace{-4mm}
\end{table}


\section{Debanding Model Development}
\label{sec:dev}

\subsection{Architecture and Training}
\label{ssec:train}

\begin{figure*}[t!]
    \begin{minipage}[b]{0.33\linewidth}
        \centering
        \centerline{\includegraphics[width=5.8cm]{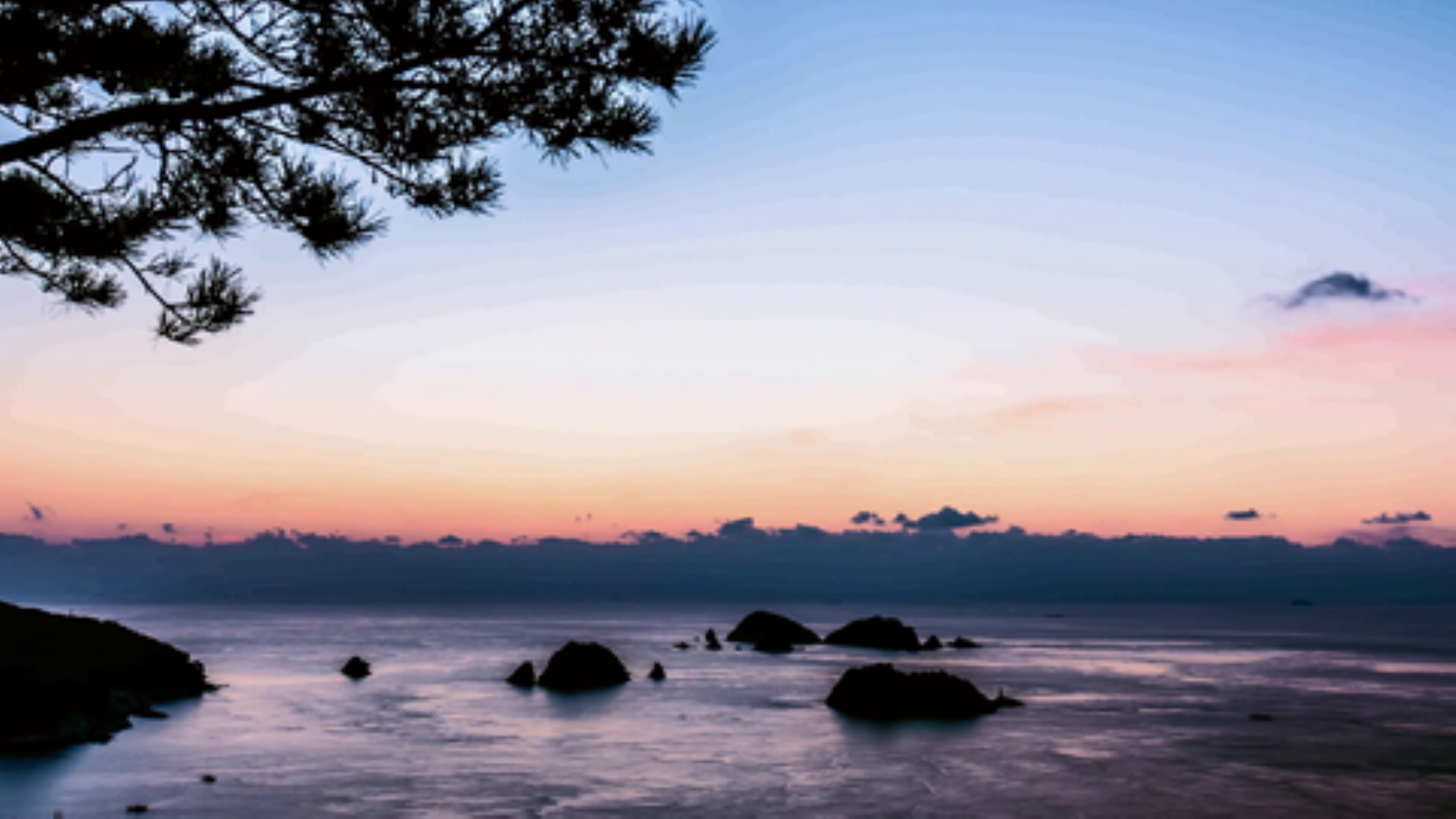}}
        \centerline{(a) Banded FHD image}\medskip
    \end{minipage}
    \hfill
    \begin{minipage}[b]{0.33\linewidth}
        \centering
        \centerline{\includegraphics[width=5.8cm]{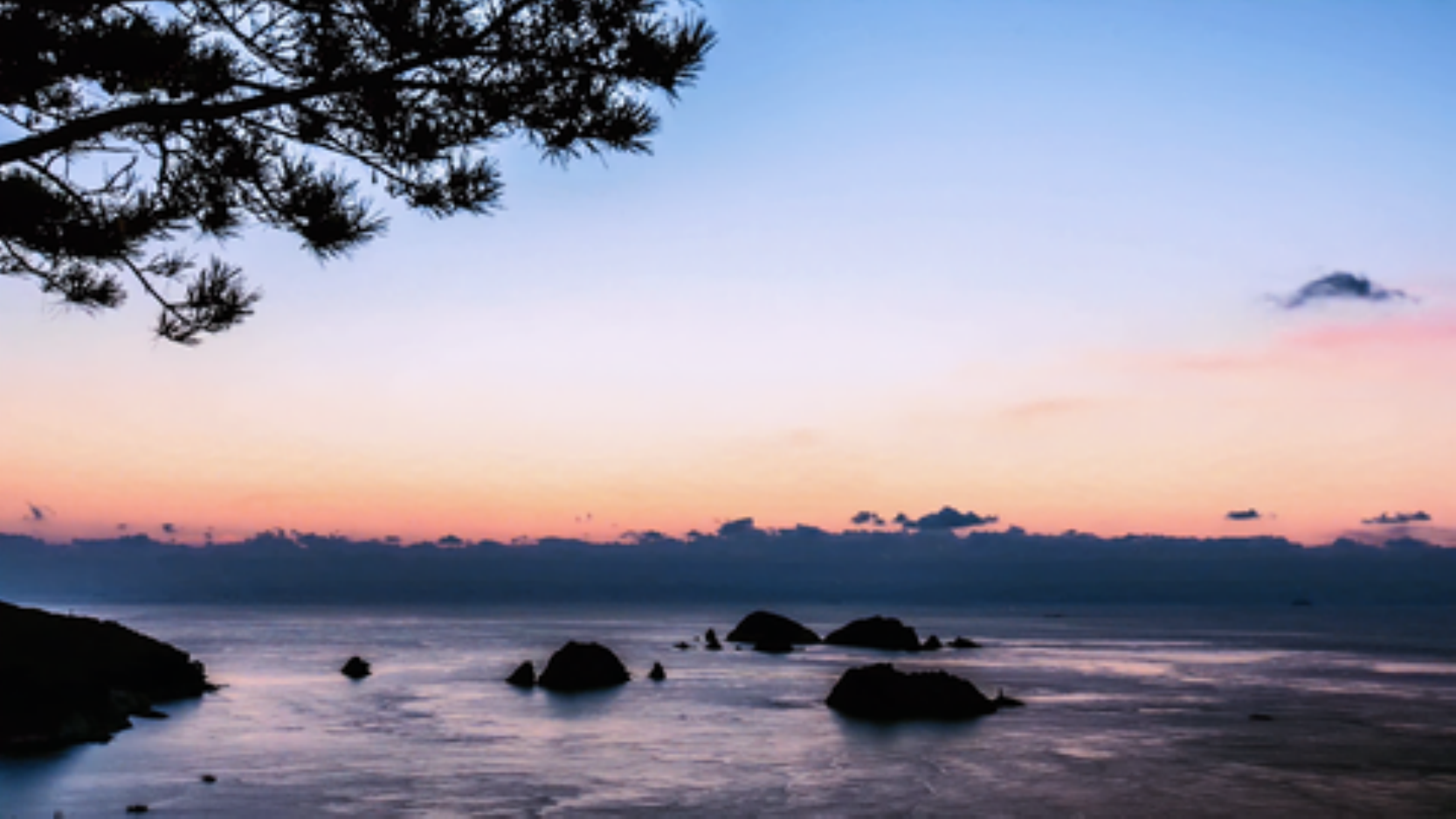}}
        \centerline{(b) deepDeband-F FHD image}\medskip
    \end{minipage}
    \hfill
    \begin{minipage}[b]{0.33\linewidth}
        \centering
        \centerline{\includegraphics[width=5.8cm]{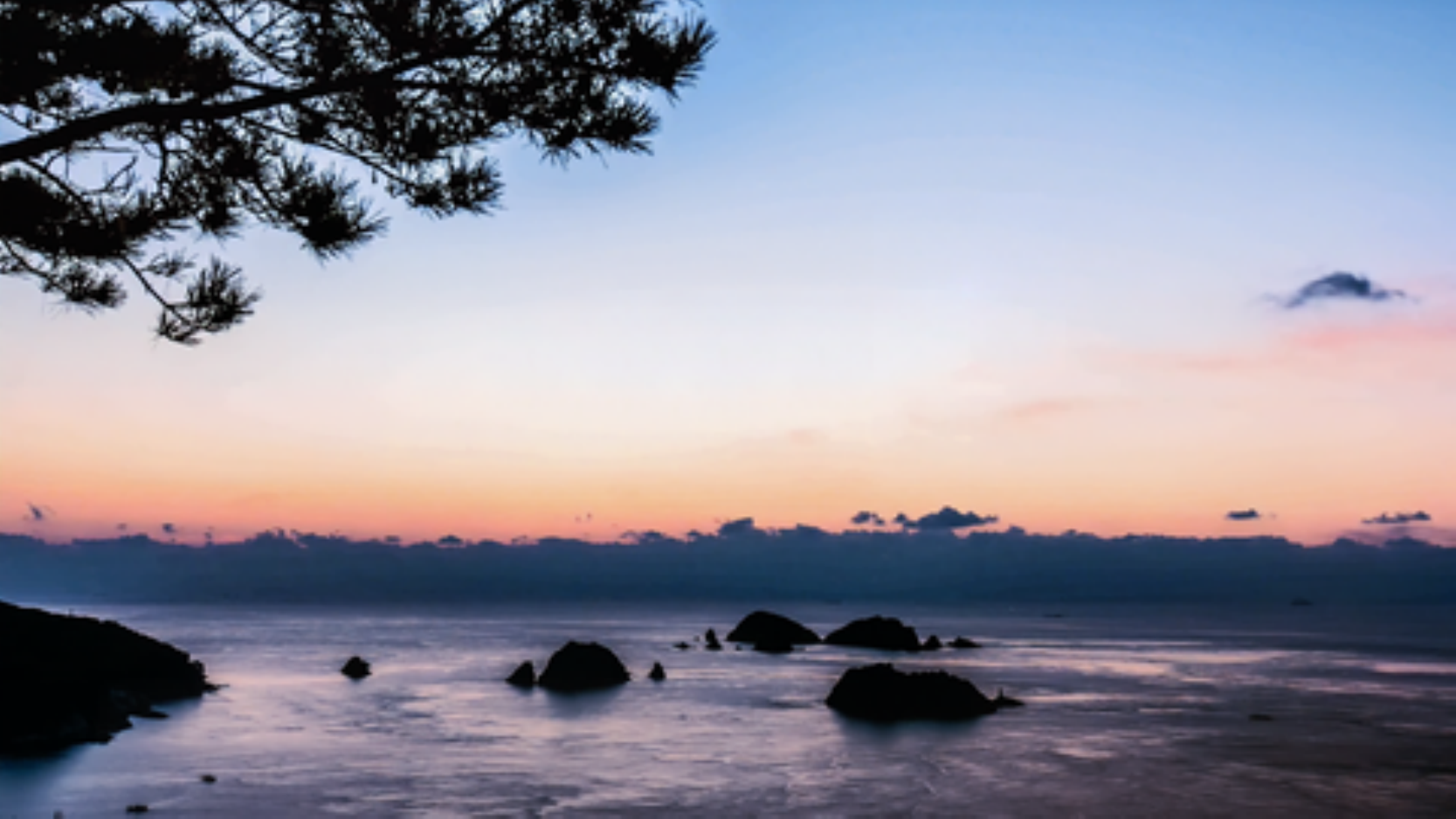}}
        \centerline{(c) deepDeband-W FHD image}\medskip
    \end{minipage}
    \vspace{-4mm}
    \caption{Banding in banded FHD image (a) is significantly reduced in deepDeband-F (b) and deepDeband-W (c) FHD images.}
    \label{fig:deband}
    \vspace{-2mm}
\end{figure*}

We use the conditional Generative Adversarial Network (cGAN) Pix2Pix \cite{pix2pix} as the basis of our deep learning model. Pix2Pix has been successfully used in other image restoration tasks, such as denoising \cite{pix2pix_noise}, deblurring \cite{pix2pix_blur}, and dehazing \cite{pix2pix_haze}, besides being successfully applied to a wide range of other image-to-image translation problems, making it promising for debanding. Compared to other networks with similar architectures, Pix2Pix has lower complexity and its adversarial objective is ideal for capturing the nature of banding artifacts and successfully removing them. 

Pix2Pix contains a generator and a discriminator \cite{pix2pix}. During training, the model is given corresponding pairs of banded and pristine 256$\times$256 patches. The generator takes the banded patches and produces debanded patches as output. The discriminator differentiates the pristine patches from the generated ones. Both parts are trained together, with the generator and discriminator trying to minimize and maximize the loss function, respectively \cite{pix2pix}. During evaluation, the generator is given a banded image and produces its debanded version. We train our debanding model, called deep debanding network (deepDeband), by using the 30,988 pairs of image patches contained in the training set of the dataset constructed in Section \ref{sec:data} (Table \ref{tab:dataset}). We also try different batch sizes, image augmentation, and dataset compositions as discussed in Section \ref{ssec:val}. All hyperparameters other than batch size are the same as the default in Pix2Pix. 

\subsection{Image-Level Application}
\label{ssec:app}

Although the model is trained on 256$\times$256 patches, we apply it on 1920$\times$1080 FHD images and use two techniques to do so. Our first method is to directly apply the model on the entire FHD image. Since the Pix2Pix generator is fully convolutional \cite{pix2pix}, deepDeband can be used to deband images of any size, though here we focus on FHD images. As the Pix2Pix generator expects an input whose width and height dimensions are divisible by 256, we first pad the input FHD image to 2048$\times$1280 through mirroring. We then crop the debanded image returned by the model back to 1920$\times$1080. We will refer to this version of the model as deepDeband Full image (deepDeband-F). Attempts at padding the input with black or white result in poor performance with unwanted textures, likely as the solid padding is unnatural visual content.

Our second method operates at the patch-level. First, we pad the input FHD image to 2048$\times$1280 and then extract overlapping 256$\times$256 patches with a stride of 128. Each patch is then debanded by the model. Next, we merge the debanded patches for each pixel $p$ in the output as 
\begin{equation}
    \centering
    {p = \frac{\sum_{i=1}^{n} w_ip_i}{\sum_{i=1}^{n} w_i}
	\label{eq:weight_merge}}
\end{equation}

\noindent where $p_i$ is the pixel value from patch $i$, and $w_i$ is the reciprocal of the distance from the pixel $p$ to the centre of patch $i$. Patches $1, 2, ..., n$ are the patches which contain pixel $p$, and $n \in \{1, 2, 4\}$ based on the location of $p$. This weighted merging approach allows us to make the best use of the overlapping patches to produce a smooth spatial transition across the image. Finally, we crop the output to 1920$\times$1080. We will refer to this version of the model as deepDeband Weighted merge (deepDeband-W). 
Fig. \ref{fig:deband} gives an example of the results of deepDeband, where it can be seen that banding in the smooth regions is greatly removed while sharp textures are well-preserved. 

\subsection{Tuning and Validation}
\label{ssec:val}

\begin{table}[t!]
    \small
    \caption{Mean validation scores of deepDeband-F for different patch sizes. Optimal values are in bold.}
    \vspace{-2mm}
    \centering
    \begin{tabular}{ c|c|c } 
    \hline
    Dataset & DBI $\downarrow$ \cite{dbi} & BBAND $\downarrow$ \cite{bband} \\ \hline \hline
    256 $\times$ 256 patches & \textbf{0.2028} & \textbf{0.1803} \\ \hline
    572 $\times$ 572 patches & 0.3495 & 0.1896 \\ \hline
    1920 $\times$ 1080 FHD images & 0.2763 & 0.1870 \\ \hline
    \end{tabular}
    \label{tab:val}
\end{table}

\begin{table}[t!]
    \small
    \caption{Mean validation scores of deepDeband-F for different dataset compositions. Optimal values are in bold.}
    \vspace{-2mm}
    \centering
    \begin{tabular}{ c|c|c } 
    \hline
    Dataset & DBI $\downarrow$ \cite{dbi} & BBAND $\downarrow$ \cite{bband} \\ \hline \hline
    Banded patches, no flipping & \textbf{0.2028} & \textbf{0.1803} \\ \hline
    Banded patches, flipping & 0.2037 & 0.1848 \\ \hline
    All patches, no flipping & 0.3114 & 0.1926 \\ \hline
    All patches, flipping & 0.3240 & 0.1924 \\ \hline
    \end{tabular}
    \label{tab:val_2}
    \vspace{-4mm}
\end{table}

Since losses reported by GANs are not meaningful due to their adversarial nature \cite{gans}, we apply DBI \cite{dbi} and BBAND \cite{bband} as evaluation metrics on the validation set of our dataset. Both DBI and BBAND are no-reference banding assessment indices, making them an ideal choice for this task. For both DBI and BBAND, a smaller value indicates less banding. To explore the optimal training conditions for our deep learning models, we create additional datasets and use the deepDeband-F model in the subsequent analysis.


First, to ascertain the optimal patch size for training, we train the deepDeband model on banded (and their corresponding pristine) patches of size 256$\times$256 and 572$\times$572, and on 1920$\times$1080 FHD images at a fixed batch size of 8. Table \ref{tab:val} shows the outcome of this experiment in terms of DBI and BBAND, where it can be seen that using the smaller 256$\times$256 patches leads to superior performance, even when applied directly at the image-level, as in deepDeband-F. This is likely because using smaller patch sizes gives more training examples, a key requirement for deep learning.


Next, we investigate different dataset compositions, using patches of size 256$\times$256 and batch size 8. In addition to the dataset of only banded patches from Section \ref{sec:data}, we use datasets containing patches of all types of visual content (banded and non-banded). We also try data augmentation by horizontally flipping the patches. Table \ref{tab:val_2} summarizes the results in terms of DBI and BBAND, where it can be seen that including all types of patches and incorporating horizontal flipping leads to poor performance which visibly manifests as undesirable textures introduced in the output images. Based on results shown in Tables \ref{tab:val} and \ref{tab:val_2}, we finalize the dataset of only banded 256$\times$256 patches, discussed in Section \ref{sec:data}, for model training.

Finally, we vary the batch size between 2, 4, 8, and 16, finding that smaller batch sizes generally give better results. We train the model for 200 epochs and choose the epoch that gives the best performance on the validation set, in terms of DBI and BBAND, as our final deepDeband models. 

\section{Performance Comparison}
\label{sec:comp}

To the best of our knowledge, deepDeband is so far the only deep learning based debanding method. Therefore, we compare its performance against three recent state-of-the-art knowledge-driven debanding methods, including FFmpeg's deband filter \cite{ffmpeg}, AdaDeband \cite{adadeband}, and FCDR \cite{fcdr}, on the 310 FHD PNG images in the test set of our dataset (Table \ref{tab:dataset}). We use default parameters for FFmpeg. We test the following four versions of AdaDeband, which was originally designed for YUV420p videos. 1) AdaDeband1: The original YUV420p version with default parameter settings; 2) AdaDeband2: The YUV420p version with parameter settings recommended by the authors for H.264 encoded frames (since images in \cite{dbi}, on which our dataset is built, were extracted from H.264 encoded videos); 3) AdaDeband3: To prevent any loss in colour due to chroma subsampling to YUV420p, we implement a YUV444p version of AdaDeband and test it at the default parameter settings; 4) AdaDeband4: The YUV444p version with parameter settings as in AdaDeband2. Since the full implementation of FCDR is not publicly available, we implement it ourselves by building upon the code provided in \cite{adadeband}. We try nine different parameter settings and choose the top two performers, which we refer to as FCDR1 and FCDR2.



For evaluation, we use both DBI \cite{dbi} (data-driven) and BBAND \cite{bband} (knowledge-driven), which to the best of our knowledge are the only publicly available banding assessment indices, making them most appropriate for our analysis. For BBAND, we choose parameter settings based on the authors' recommendations for content in our test set. As additional evaluation metrics, we use three blind image quality assessment (BIQA) methods of varying design philosophies, which are shown to be top performers in the BIQA area in \cite{iqa}. These include HOSA (opinion-aware) \cite{hosa}, dipIQ (opinion-unaware) \cite{dipiq}, and ILNIQE (opinion-unaware) \cite{ilniqe}.


\begin{table*}[t!]
    \small
    \caption{Performance comparison of debanding methods in terms of mean ($\pm$SD) evaluation metric scores on the entire test set. Optimal values are in bold. The arrow besides the metric name shows whether higher or lower mean values are better.}
    \vspace{-1mm}
    \centering
    \begin{tabular}{ c|c|c|c|c|c } 
    \hline
    Model & DBI $\downarrow$ \cite{dbi} & BBAND $\downarrow$ \cite{bband} & dipIQ $\uparrow$ \cite{dipiq} & HOSA $\downarrow$ \cite{hosa} & ILNIQE $\downarrow$ \cite{ilniqe} \\ \hline \hline
    Banded Images & 0.4059 ($\pm$0.3599) & 0.3830 ($\pm$0.2214) & -6.5261 ($\pm$2.5607) & 34.8581 ($\pm$11.8420) & 29.7630 ($\pm$18.0335) \\ \hline \hline
    FFmpeg \cite{ffmpeg} & 0.2240 ($\pm$0.2439) & 0.1523 ($\pm$0.0743) & -5.8885 ($\pm$2.6462) & 34.6408 ($\pm$\textbf{9.8343}) & 30.0330 ($\pm$17.8628) \\ \hline
    AdaDeband1 \cite{adadeband} & 0.3414 ($\pm$0.3327) & 0.2085 ($\pm$0.1218) & -6.3723 ($\pm$2.5196) & 34.8560 ($\pm$11.1266) & 28.3333 ($\pm$13.4799) \\ \hline
    AdaDeband2 \cite{adadeband} & 0.3358 ($\pm$0.3313) & 0.2060 ($\pm$0.1147) & -6.3735 ($\pm$2.5184) & 34.9581 ($\pm$11.1495) & 28.4328 ($\pm$13.6613) \\ \hline
    AdaDeband3 \cite{adadeband} & 0.3374 ($\pm$0.3321) & 0.2163 ($\pm$0.1266) & -6.4375 ($\pm$2.5854) & 35.1767 ($\pm$11.2836) & 28.1319 ($\pm$13.8442) \\ \hline
    AdaDeband4 \cite{adadeband} & 0.3328 ($\pm$0.3306) & 0.2135 ($\pm$0.1180) & -6.4370 ($\pm$2.5843) & 35.1811 ($\pm$11.3185) & 28.2034 ($\pm$14.1324) \\ \hline
    FCDR1 \cite{fcdr} & 0.3980 ($\pm$0.3599) & 0.3468 ($\pm$0.2034) & \textbf{-4.9563} ($\pm$2.4227) & 35.4358 ($\pm$11.8459) & 29.0935 ($\pm$17.6923) \\ \hline
    FCDR2 \cite{fcdr} & 0.3813 ($\pm$0.3558) & 0.3538 ($\pm$0.2043) & -5.3961 ($\pm$\textbf{2.1078}) & 35.7727 ($\pm$11.6396) & 28.8689 ($\pm$17.7932) \\ \hline \hline
    deepDeband-F & 0.2026 ($\pm$0.2369) & \textbf{0.1518 ($\pm$0.0737)} & -5.3110 ($\pm$2.4216) & \textbf{32.8358} ($\pm$9.9899) & \textbf{25.4175 ($\pm$10.5596)} \\ \hline
    deepDeband-W & \textbf{0.1774 ($\pm$0.1962)} & 0.1655 ($\pm$0.0823) & -5.8916 ($\pm$2.4793) & 34.6312 ($\pm$10.7856) & 26.1697 ($\pm$10.8724)\\ \hline
    \end{tabular}
    \label{tab:eval}
\end{table*}

\begin{figure*}[t!]
\begin{minipage}[b]{0.37\linewidth}
  \centering
  \centerline{\includegraphics[width=6cm]{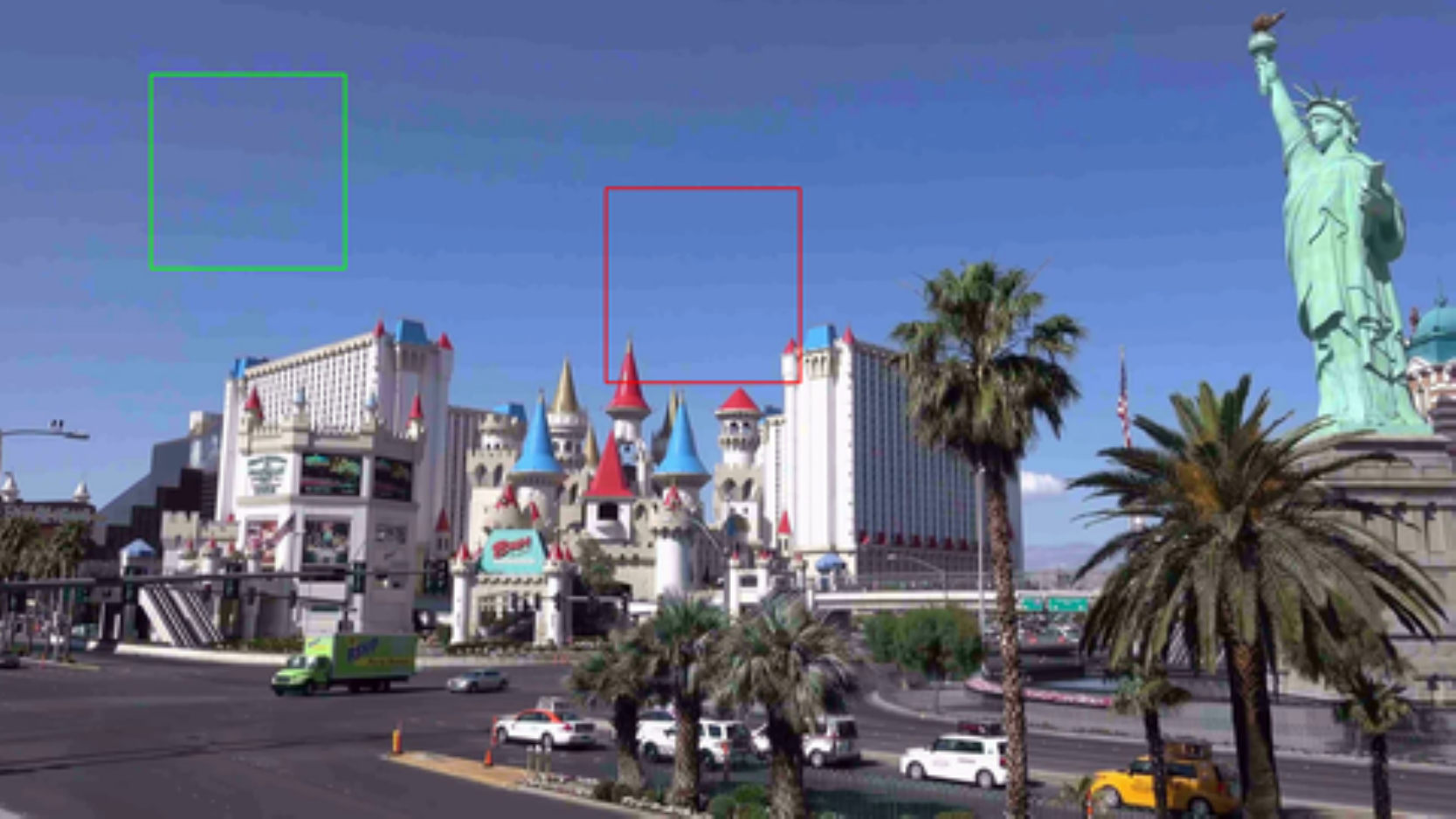}}
  \centerline{}\medskip
\end{minipage} \hfill
\begin{minipage}[b]{0.1\linewidth}
  \centering
  \centerline{\includegraphics[width=1.6cm]{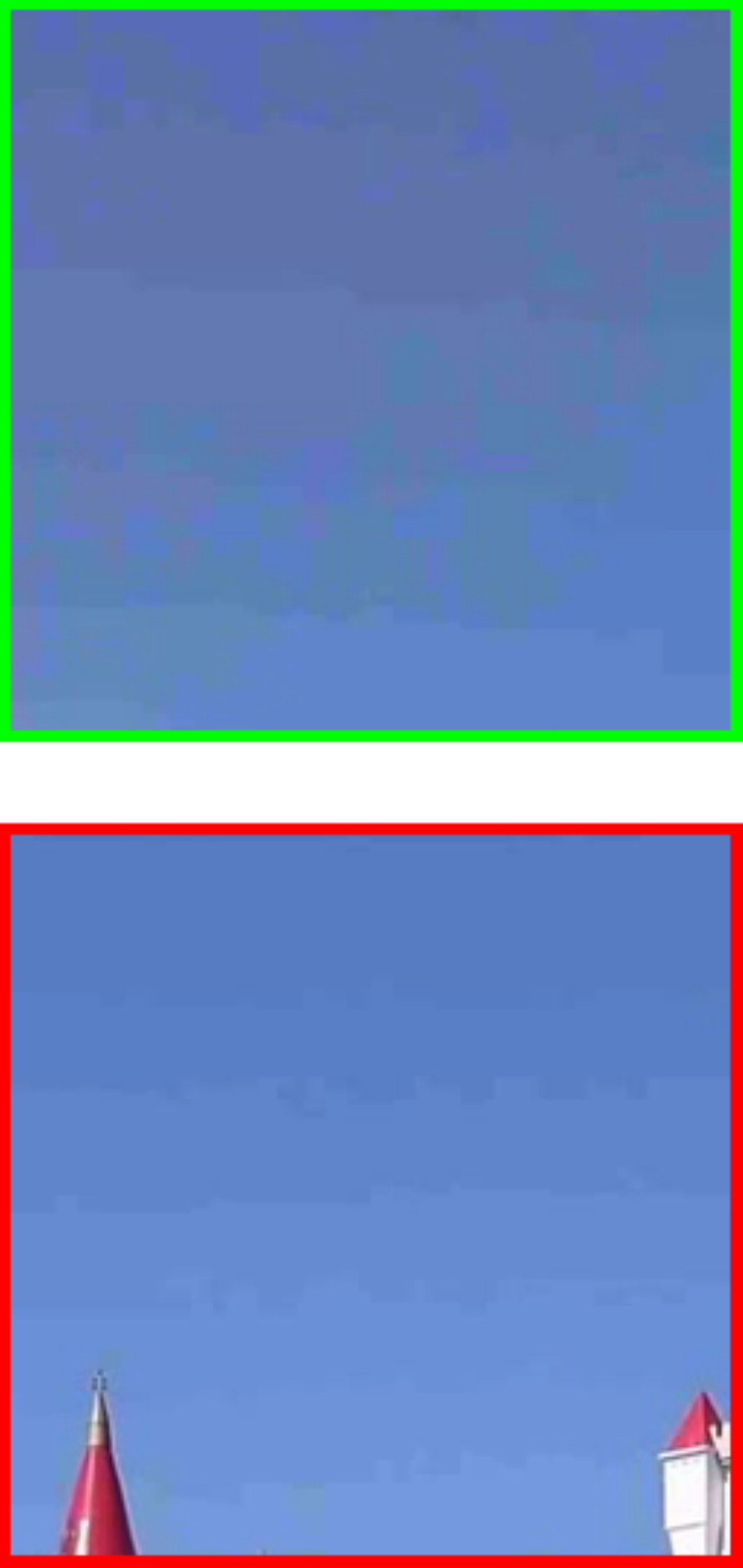}}
  \centerline{}\medskip
\end{minipage} \hfill 
\begin{minipage}[b]{0.1\linewidth}
  \centering
  \centerline{\includegraphics[width=1.6cm]{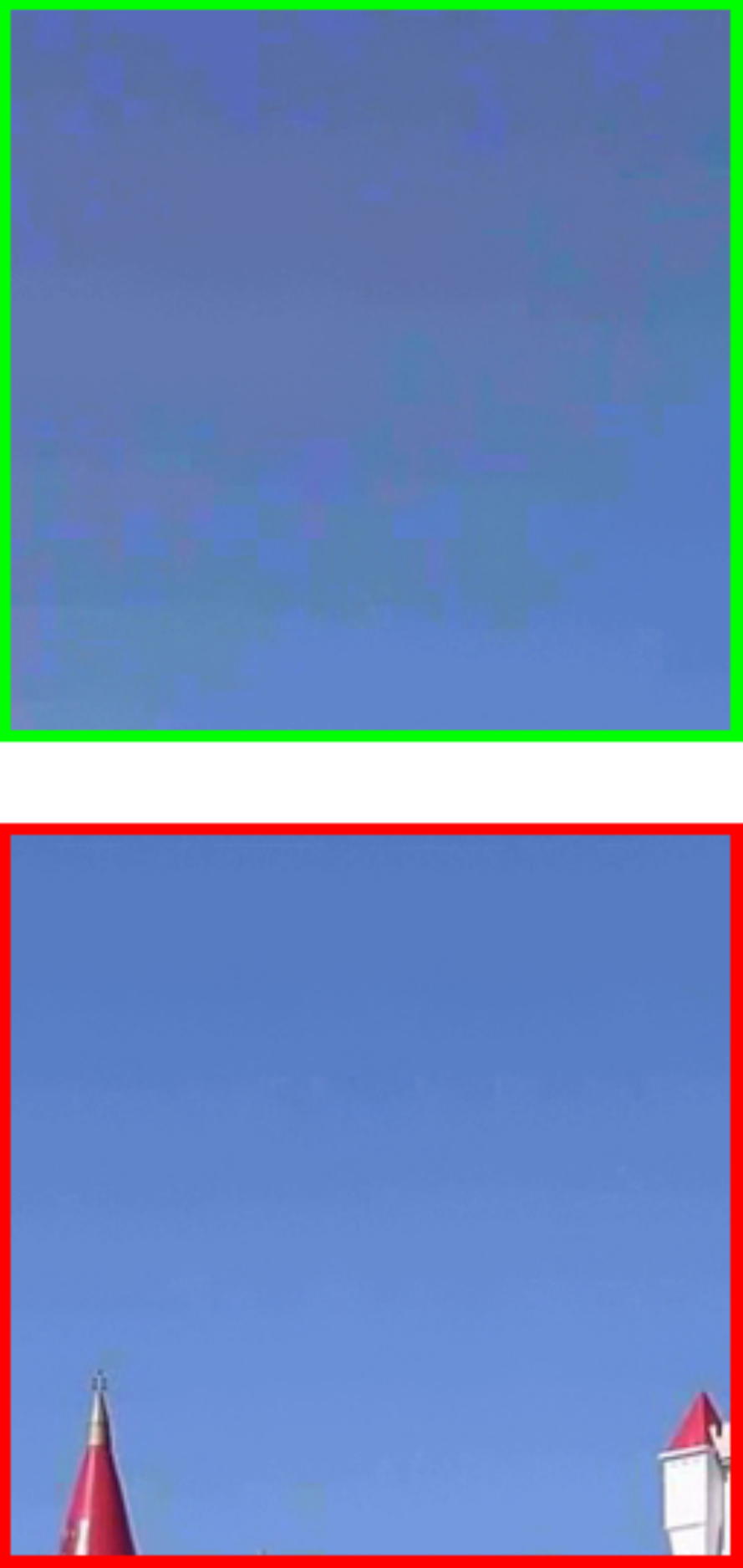}}
  \centerline{}\medskip
\end{minipage} \hfill 
\begin{minipage}[b]{0.1\linewidth}
  \centering
  \centerline{\includegraphics[width=1.6cm]{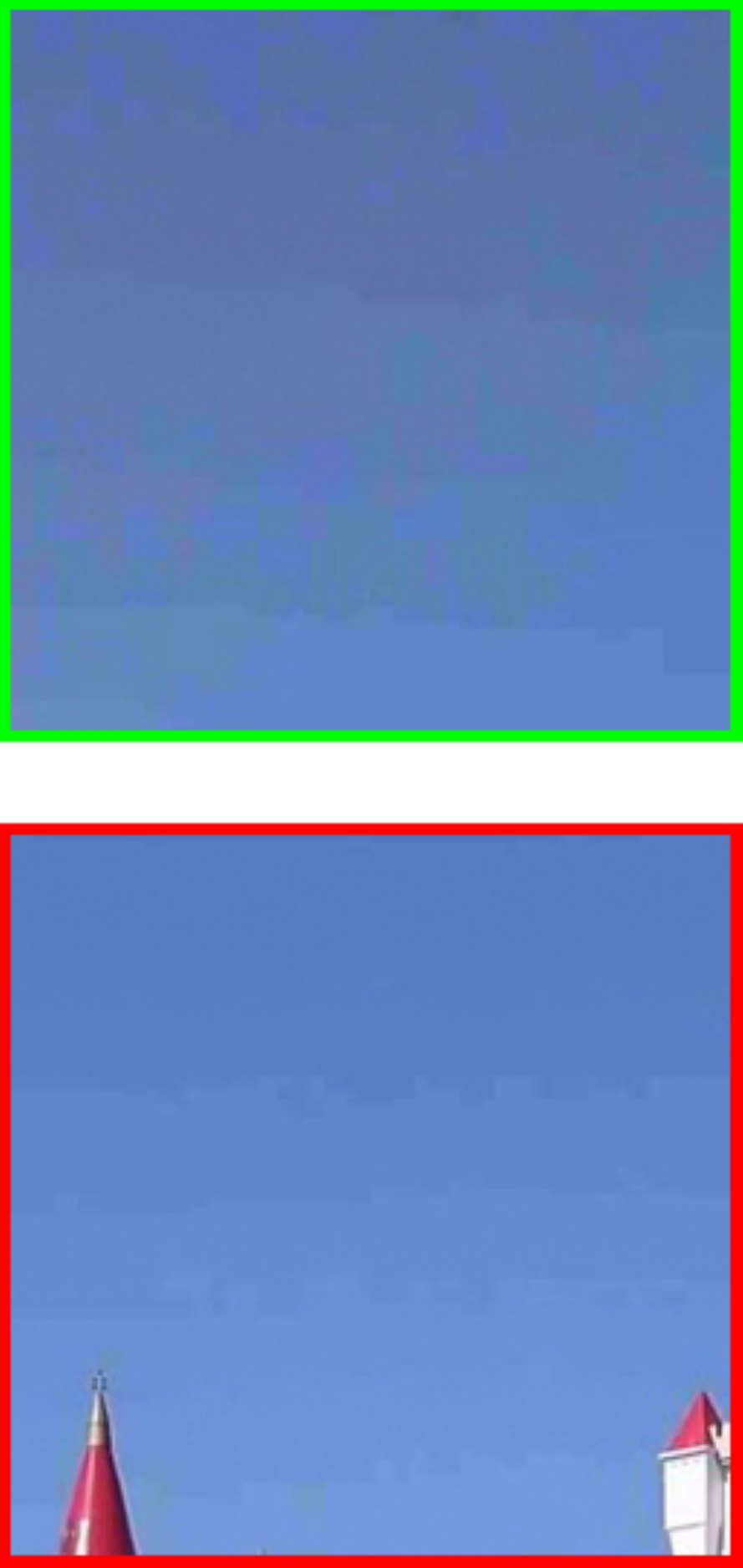}}
  \centerline{}\medskip
\end{minipage} \hfill 
\begin{minipage}[b]{0.1\linewidth}
  \centering
  \centerline{\includegraphics[width=1.6cm]{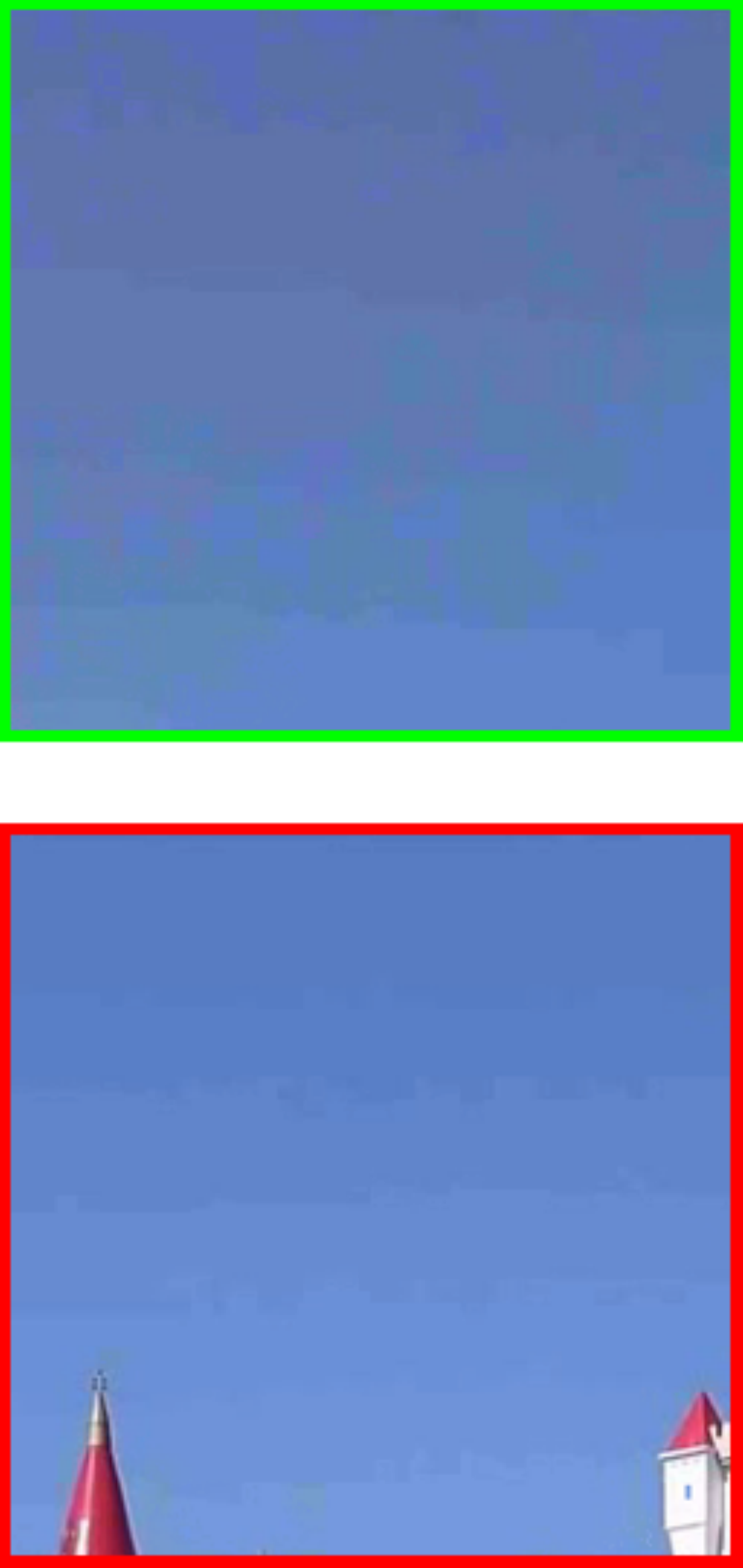}}
  \centerline{}\medskip
\end{minipage} \hfill 
\begin{minipage}[b]{0.1\linewidth}
  \centering
  \centerline{\includegraphics[width=1.6cm]{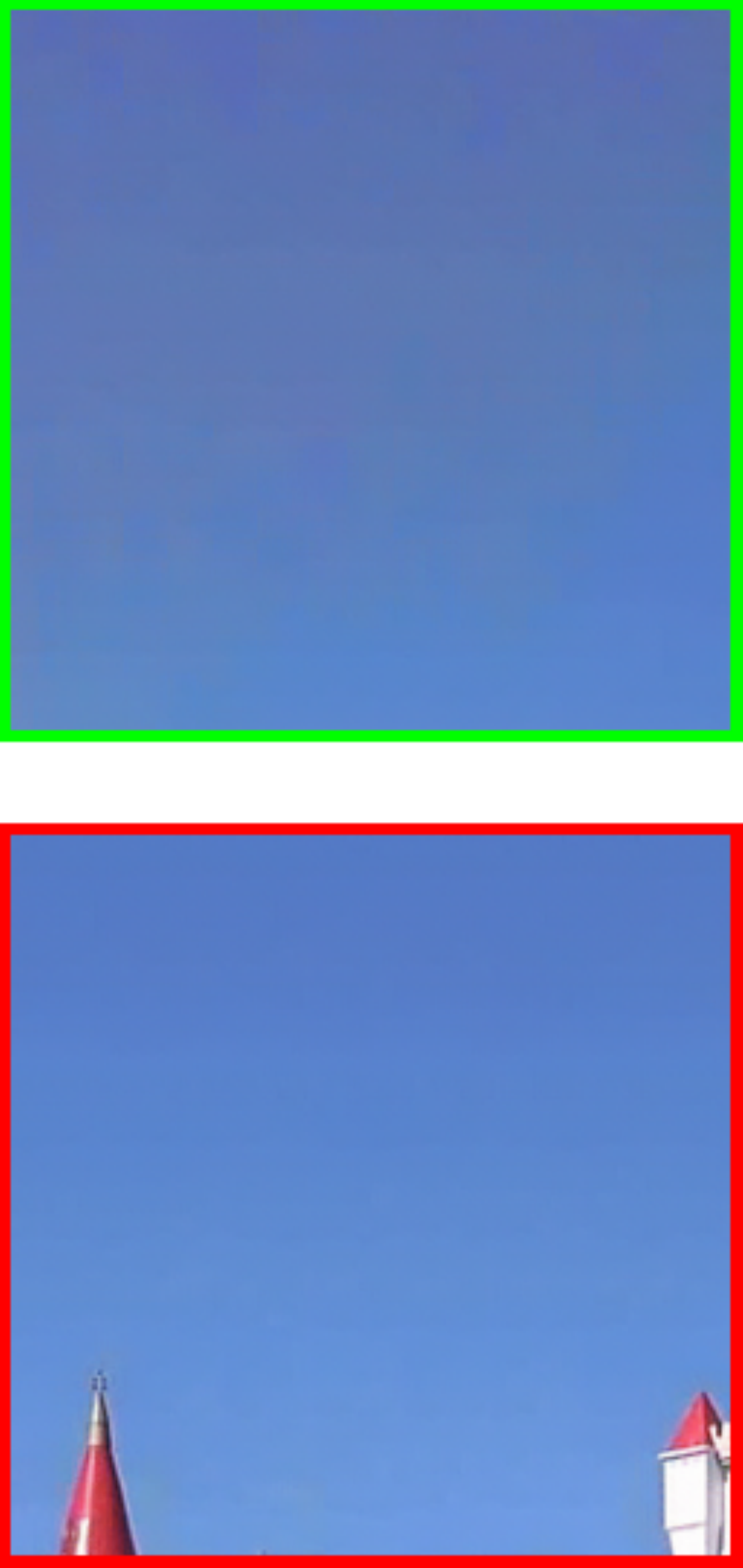}}
  \centerline{}\medskip
\end{minipage}
\begin{minipage}[b]{0.1\linewidth}
  \centering
  \centerline{\includegraphics[width=1.6cm]{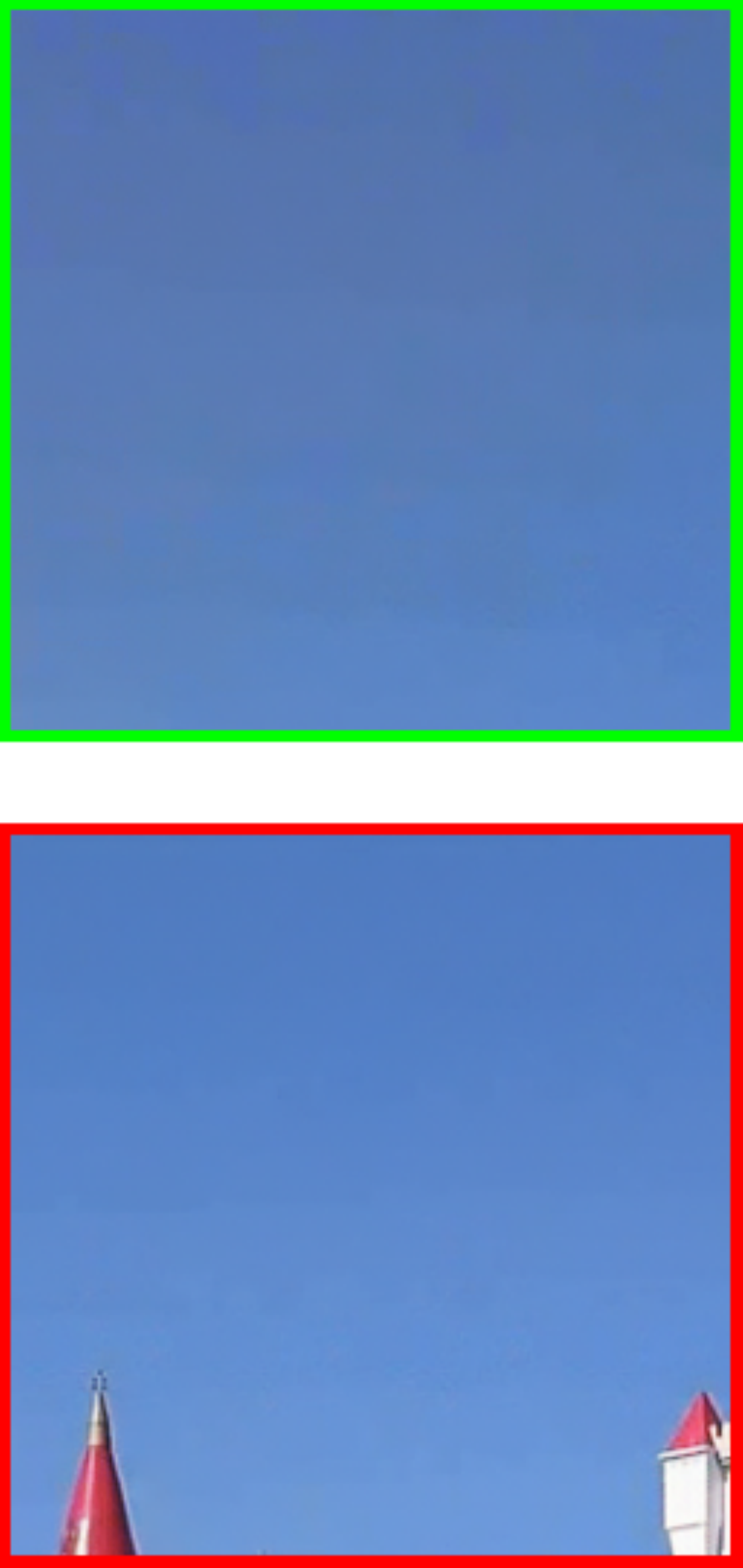}}
  \centerline{}\medskip
\end{minipage}
\vspace{-4mm}
\begin{minipage}[b]{0.37\linewidth}
  \centering
  \centerline{\includegraphics[width=6cm]{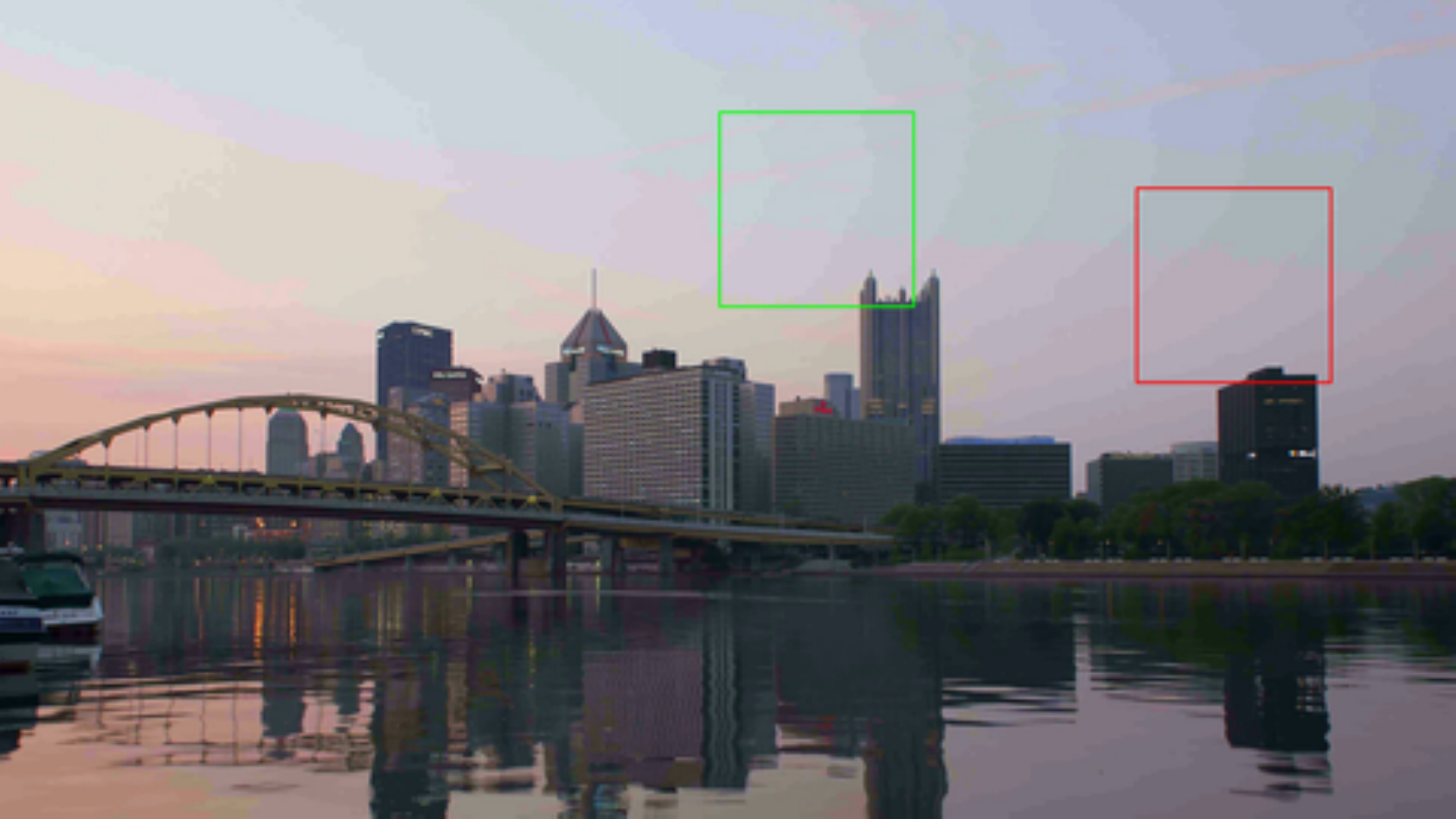}}
  \centerline{Banded FHD Image}\medskip
\end{minipage} \hfill
\begin{minipage}[b]{0.1\linewidth}
  \centering
  \centerline{\includegraphics[width=1.6cm]{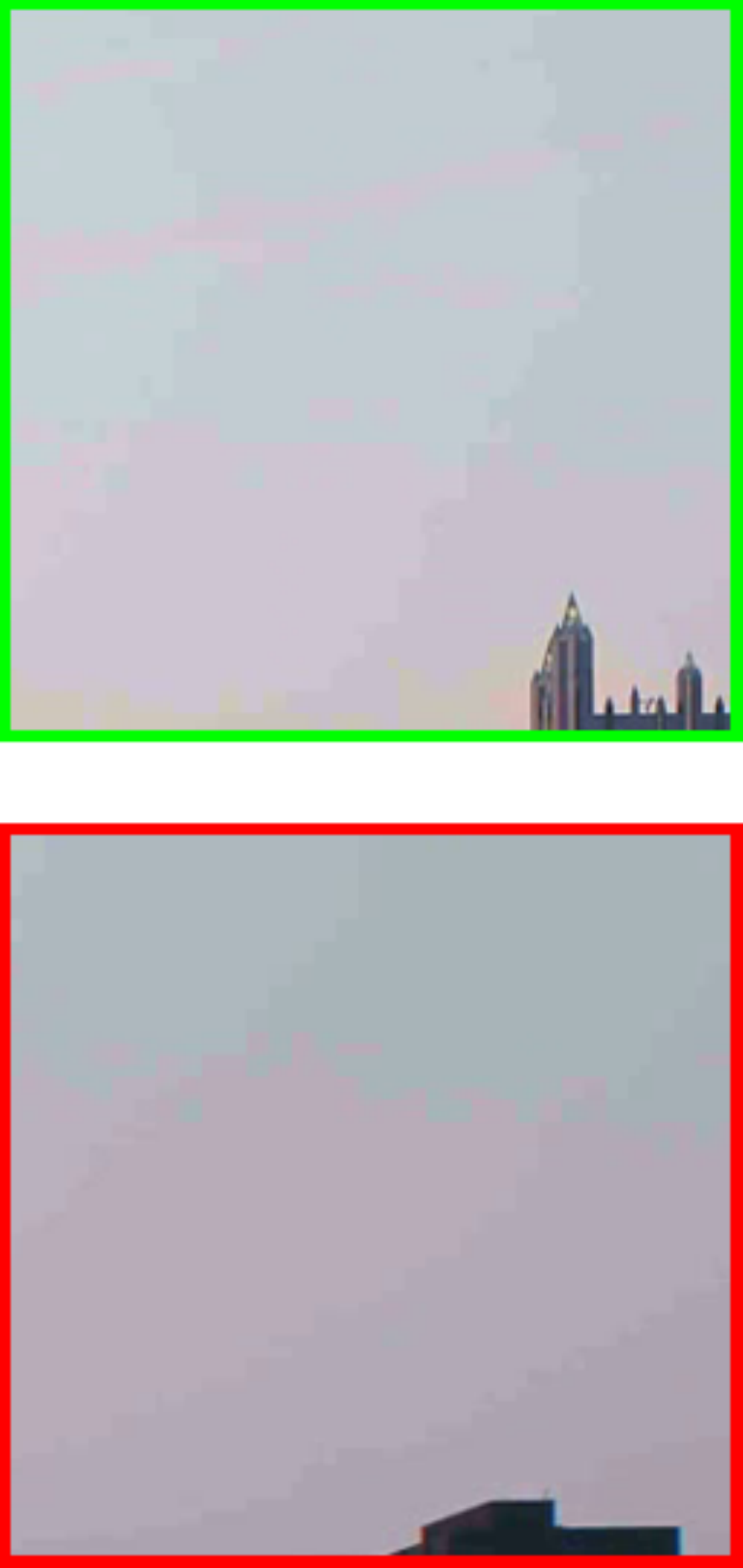}}
  \centerline{(a)}\medskip
\end{minipage} \hfill 
\begin{minipage}[b]{0.1\linewidth}
  \centering
  \centerline{\includegraphics[width=1.6cm]{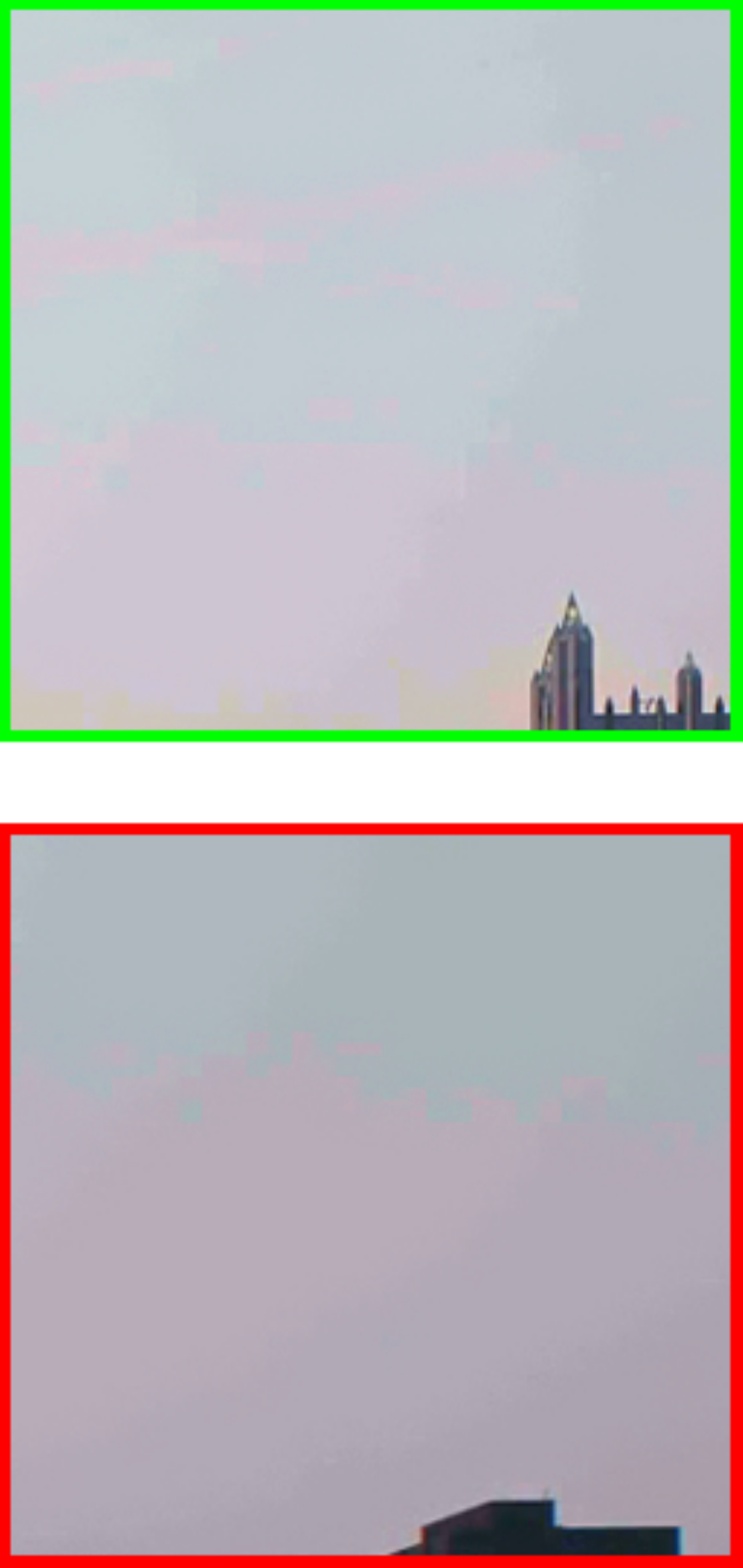}}
  \centerline{(b)}\medskip
\end{minipage} \hfill 
\begin{minipage}[b]{0.1\linewidth}
  \centering
  \centerline{\includegraphics[width=1.6cm]{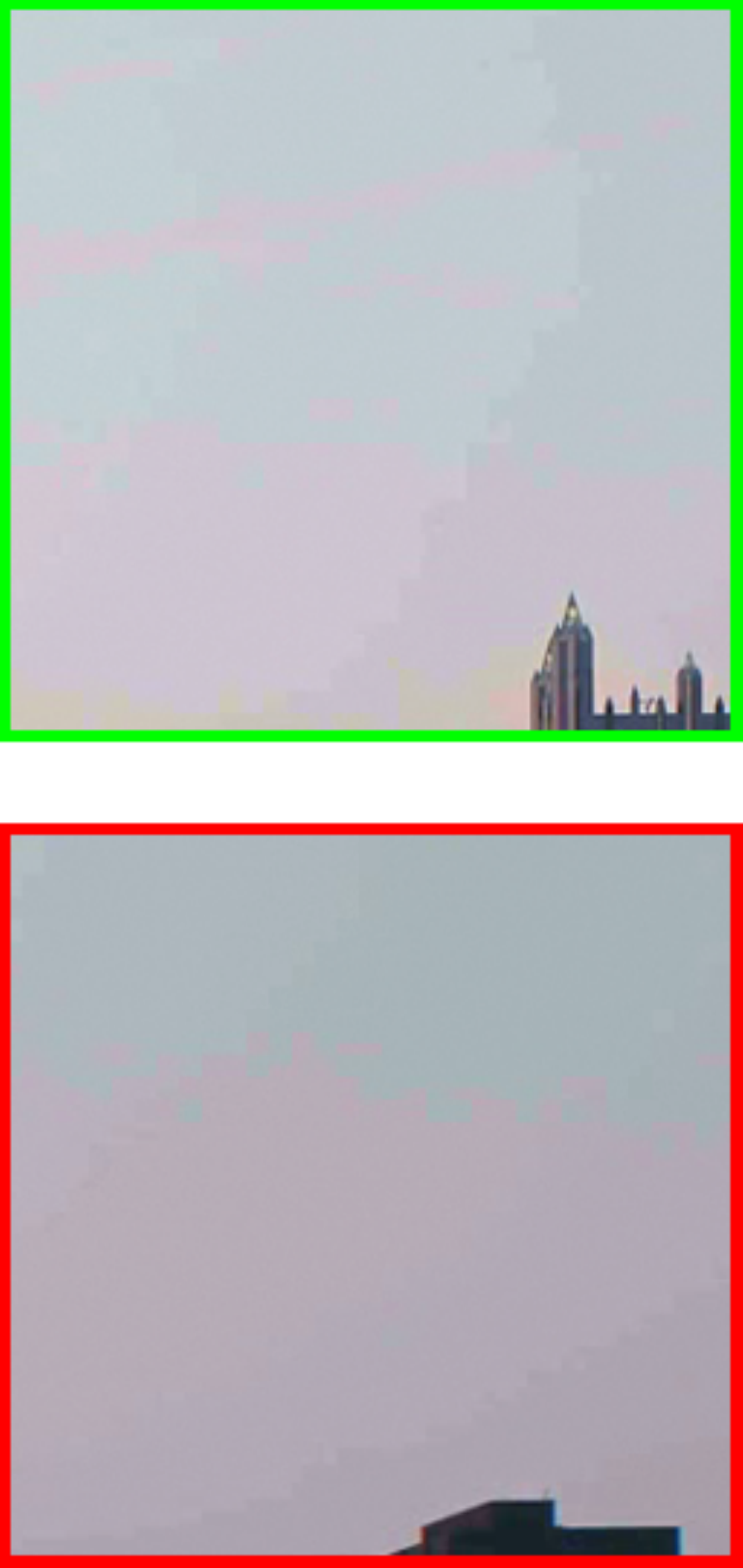}}
  \centerline{(c)}\medskip
\end{minipage} \hfill 
\begin{minipage}[b]{0.1\linewidth}
  \centering
  \centerline{\includegraphics[width=1.6cm]{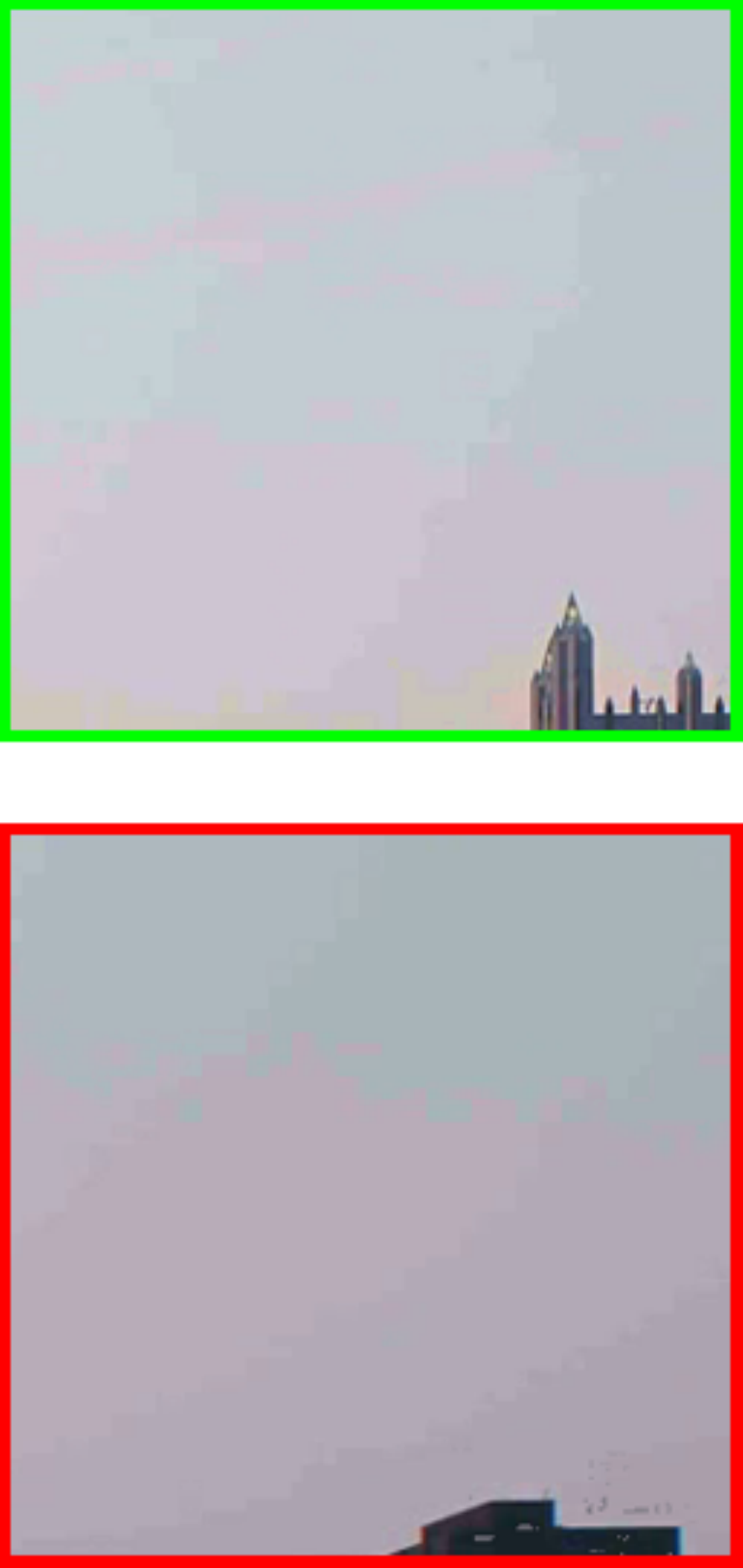}}
  \centerline{(d)}\medskip
\end{minipage} \hfill 
\begin{minipage}[b]{0.1\linewidth}
  \centering
  \centerline{\includegraphics[width=1.6cm]{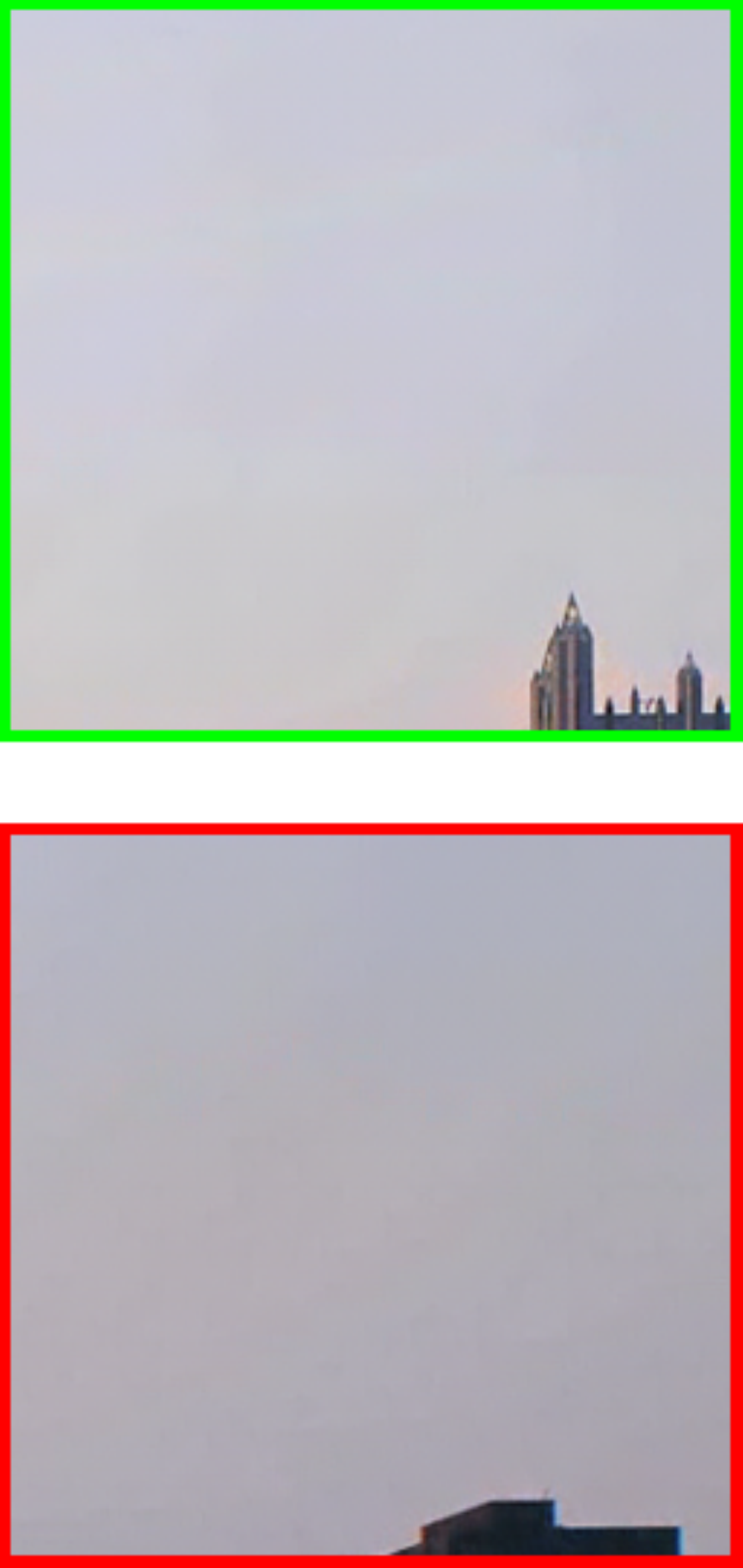}}
  \centerline{(e)}\medskip
\end{minipage}
\begin{minipage}[b]{0.1\linewidth}
  \centering
  \centerline{\includegraphics[width=1.6cm]{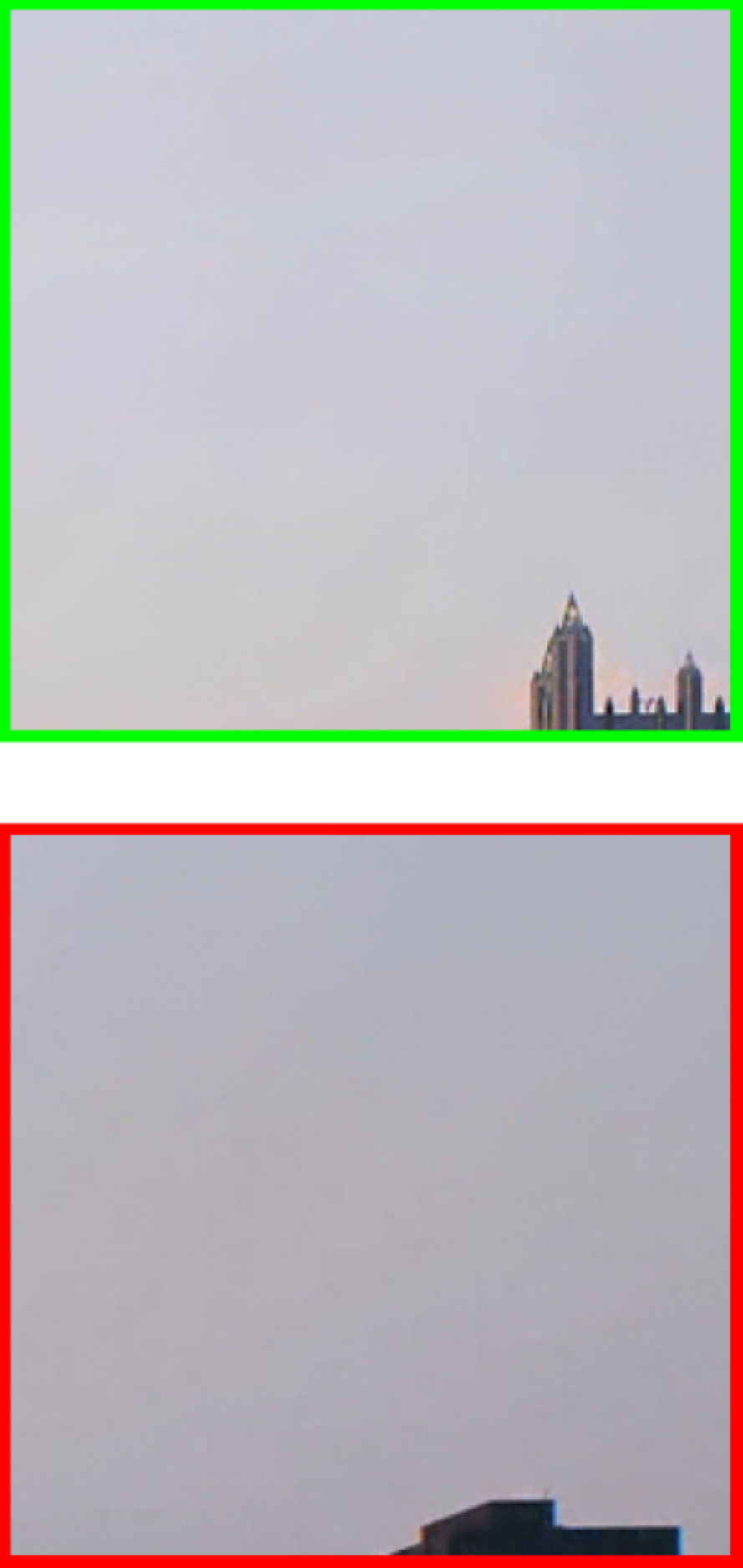}}
  \centerline{(f)}\medskip
\end{minipage}
\vspace{2mm}
\caption{Visual comparison of banded (a), and debanded content using FFmpeg (b), AdaDeband4 (c), FCDR2 (d), deepDeband-F (e), and deepDeband-W (f).}
\label{fig:visual}
\vspace{-4mm}
\end{figure*}

\begin{table}[t!]
    \small
    \caption{Execution time comparison. Optimal values in bold.}
    \vspace{-1mm}
    \centering
    \begin{tabular}{ c|c } 
    \hline
    Model & Time (seconds) $\downarrow$ \\ \hline \hline
    FFmpeg \cite{ffmpeg} & \textbf{1.2907} \\ \hline
    AdaDeband1 \cite{adadeband} & 10.6903 \\ \hline
    AdaDeband2 \cite{adadeband} & 12.7411 \\ \hline
    AdaDeband3 \cite{adadeband} & 10.9448 \\ \hline
    AdaDeband4 \cite{adadeband} & 12.8803 \\ \hline
    FCDR1 \cite{fcdr} & 23.4890 \\ \hline
    FCDR2 \cite{fcdr} & 38.7096 \\ \hline
    deepDeband-F & 10.0121 \\ \hline
    deepDeband-W & 154.8274 \\ \hline
    \end{tabular}
    \label{tab:eval_speed}
    \vspace{-4mm}
\end{table}

Table \ref{tab:eval} shows the performance of debanding methods under test in terms of mean evaluation metric scores along with their standard deviation (SD) in parentheses. It also includes scores for the original banded images as a baseline. Notably, deepDeband-F surpasses all existing models for all metrics except FCDR1 in terms of dipIQ, and its DBI score is second only to that of deepDeband-W. Similarly, deepDeband-W outperforms all existing models except FFmpeg in terms of BBAND, and FCDR1 and FCDR2 in terms of dipIQ, making it as competitive as deepDeband-F. Furthermore, deepDeband-F and deepDeband-W perform significantly better than FCDR1 and FCDR2 for all metrics other than dipIQ, especially in terms of DBI and BBAND. We also note that deepDeband-F and deepDeband-W generally give comparatively lower standard deviation values, suggesting that they are relatively more reliable than other debanding methods. We also present a visual comparison of the results in Fig. \ref{fig:visual}, as the ultimate goal is to reduce human perception of banding. It can be seen that although FFmpeg, AdaDeband4, and FCDR2 reduce the visibility of banding, the banding contours are still visible, which are hardly discerned in images produced by deepDeband methods.



Finally, we evaluate the computational complexity of different methods in terms of average execution time to deband an FHD image on a machine with a 2.70GHz Intel Core i7-7500U processor and 8GB of RAM. Table \ref{tab:eval_speed} shows the results for this experiment, where it can be seen that deepDeband-F is faster than all other methods except FFmpeg, while deepDeband-W is the slowest, which can be attributed to the weighted merge approach that it takes. However, both of our models can be accelerated by GPUs during evaluation.


\section{Conclusion}
\label{sec:concl}

We propose the first deep learning based model of its kind for removing banding artifacts from images, deepDeband. It possesses none of the disadvantages of existing knowledge-driven methods, such as the need to carefully fine-tune parameters, and can be applied to images of any size. Extensive performance evaluation shows that deepDeband outperforms all existing debanding methods, both quantitatively and visually. To train deepDeband, we create a dataset of 51,490 pairs of image patches, comprised of corresponding banded and pristine patches. The model builds the foundation to support future work in deep learning to identify and remove banding artifacts from visual content.


\bibliographystyle{IEEEbib}
\bibliography{IEEEabrv,refs}

\end{document}